# $^{59}$Co-NMR Knight Shift of Aligned Crystals and Polycrystalline Samples of Superconducting $Na_{0.3}CoO_2 \cdot 1.3H_2O$

Yoshiaki KOBAYASHI, Hidekazu WATANABE, Mai YOKOI, Taketo MOYOSHI, Yoshihiko MORI and Masatoshi SATO

Department of Physics, Division of Material Science, Nagoya University, Furo-cho, Chikusa-ku, Nagoya 464-8602

Temperature ($T$) dependence of $^{59}$Co-NMR Knight shifts $K$ of $Na_{0.3}CoO_2.1.3H_2O$ has been studied, where samples of randomly oriented powder and aligned crystals have been used for the applied magnetic fields $\boldsymbol{H}$ // *ab* plane and $\boldsymbol{H}$ //*c*-axis, respectively. For both directions of $\boldsymbol{H}$, the shift $K$ decreases below the superconducting transition temperature $T_c(H)$ with decreasing $T$, indicating that the superconducting electron pairs are in the singlet state. The upper critical fields $H_{c2}(T)$ determined from the $K(H)$-$T$ curves are found to be consistent with the values reported by the resistivity measurements for both directions of $\boldsymbol{H}$.

corresponding auther: M. Sato (e-mail: e43247a@nucc.cc.nagoya-u.ac.jp)

The superconductivity found by Takada *et al.*[1] in $Na_{0.3}CoO_2{\cdot}1.3H_2O$ appears in the two dimensional triangular lattice of Co atoms. Due to this structural characteristic, much attention has been paid to the mechanism of the superconductivity, because the superconducting state may be considered to be realized in geometrically frustrated electrons, as is strongly anticipated by many researchers after the well-known resonating-valence-bond (RVB) theory[2] developed in the study of high-$T_c$ Cu oxides. Another characteristic of the electron system which we are interested in, is the fact that $Co^{3+}$ and/or $Co^{4+}$ ions in various Co oxides often exhibit the spin state change.[3-5] For example, the low spin (LS, $t_{2g}^6$) ground state of the 3*d* electrons of $Co^{3+}$ in the pseudo cubic crystal field often changes to the intermediate spin (IS, $t_{2g}^5e_g^1$) or high spin (HS, $t_{2g}^4e_g^2$) state with varying temperature $T$ or with the substitution of constituent elements. Because the $e_g$-electrons of the ions in the IS- or HS-state induce the double exchange interaction between the spins in the $t_{2g}$ orbitals, the metallic and ferromagnetic (or nearly ferromagnetic) state is easily realized in many Co oxides. Then, this consideration suggests a possibility of the superconductivity induced by the interband interaction near the ferromagnetic state. In this sense, it is interesting that many theoretical groups have suggested the unconventional nature of the superconductivity with triplet Cooper pairs,[6-9] though the pairing state induced by the interband interaction may not necessarily be triplet one.[10]

In order to investigate the pairing state of the present system, we have carried out NQR- and NMR-studies and reported the $T$ dependences of the NQR longitudinal relaxation rate ($1/T_1$)[11] and the Knight shift[12] : As for the relaxation rate, we reported that a very tiny coherence peak was observed in the $1/T_1T$-$T$ curve. However, because the existence of the coherence peak has not been confirmed in the succeeding measurements carried out for other pure and impurity doped samples,[13] a possibility cannot be excluded that the superconducting order parameter does not have a fully gapped order parameter. The data of the NMR Knight shift, which we reported in the previous paper, are taken by using a polycrystalline sample for a peak in the NMR spectra, which corresponds to the crystal orientation of ***H***//(one of the local principal axes, *y* within the *ab* plane). The results clearly indicate that the spin susceptibility along the *y* direction is suppressed by the occurrence of the superconductivity, indicating that the superconducting pairs are in the singlet state, unless the spins of the Cooper pairs are strongly pinned along the *c* direction. (We think that the spins of the Cooper pairs are not pinned within the *ab* plane.)

In the present paper, we report results of Knight shift studies carried out in the field ***H***//*c* on aligned crystal samples. We also report results of detailed studies of the Knight shift on a powder sample in the field ***H***//*y*. The data show that the spin susceptibility is suppressed by the superconductivity in both directions, excluding the possibility of the triplet pair state of the present system. From the $H$-dependence of the $K(H)$-$T$ curves, the upper critical fields ($H_{c2}$)-$T$ curve has been obtained. These values are consistent with those reported by resistivity measurements[14-16] for both directions of the applied magnetic field ***H***.

Polycrystalline samples of $Na_{0.72}CoO_2$ were first prepared. Then, the superconducting powder samples of $Na_{0.3}CoO_2{\cdot}1.3H_2O$ were prepared by de-intercalating Na from the polycrystals and by intercalating $H_2O$ molecules into them.[11] A randomly oriented powder sample thus obtained was used in the NMR measurements. Single crystals of $Na_xCoO_2$ with $x$~0.72 were grown by the floating zone (FZ) method as reported in ref.16. They were crushed into platelets, for which the Na de-intercalation and the $H_2O$ intercalation were carried out to obtain superconducting samples of $Na_{0.35}CoO_2 \cdot 1.3H_2O$. Then, their $c$ axes were aligned. In the NMR measurements, $\boldsymbol{H}$ was applied parallel to the aligned $c$ axis. The X-ray diffraction patterns and $^{59}$Co-NMR spectra taken for the aligned samples indicate that they have not only the superconducting crystals with bi-layer of $H_2O$ between the $CoO_2$ planes (bi-layer phase) but also the non-superconducting crystals with monolayer of $H_2O$ between the $CoO_2$ planes (monolayer phase) and non-hydrated crystals. (For the powder sample, we have detected neither the monolayer phase nor the non-hydrated crystals)

The measurements are carried out by standard coherent pulsed NMR method. The NMR spectra were taken by recording the spin echo intensity with the applied field being changed stepwise. In the measurements of the aligned crystals, the central transition (-1/2 $\leftrightarrow$ 1/2) line was obtained by the Fourier transform (FT) NMR technique.

Figure 1(a) shows the NMR spectra observed for the randomly oriented powder sample at the fixed frequency of 64.000 MHz in a magnetic field range of 5.4~7.0 T. The dotted curve shows the result of the fitting to the observed data (open circles) carried out by considering the anisotropic Knight shifts and the electric quadrupole (*eqQ*) interaction up to the second order, where the positions of peaks and shoulders of the NMR spectra are well explained.[12)] We define the $z$ axis along the principal direction for which the component $\nu_{zz}$ of the electric quadrupole tensor is maximum. Then, because the value of $\nu_{cc}$ obtained from the NMR spectra of aligned crystals agrees with the $\nu_{zz}$ value deduced in the above analysis of the powder spectra, the $c$ axis is found to be parallel to the $z$ axis. The peak position indicated by the arrow in the figure corresponds to the resonance field $H_{\mathrm{res},y}$ of $Na_{0.3}CoO_2 \cdot 1.3H_2O$. Here, the profiles of the peak have been taken at various temperatures $T$ with various fixed frequencies ($f$) of 16.000, 22.989, 30.044 and 47.000MHz. In Figs. 1(b) and 1(c), the results for $\boldsymbol{H}//y$ are shown for $f$ = 22.989 and 47.000MHz, respectively. The resonance field $H_{\mathrm{res},\,y}$ determined as the peak position of the profiles, is nearly $T$-independent between $T_c$ and ~40 K as shown later and begins to decrease at $T_c$ with decreasing $T$ at all frequencies studied here.

In Fig. 2(a), the values of $\Delta H_y\,(T) \equiv H_{\mathrm{res},\,y}\,(T)$- $H_{\mathrm{res},\,y}$ (4.7 K) are shown against $T$ for various fixed values of the NMR frequencies $f$. Although the shift of the resonance field $H_{\mathrm{res},\,y}$ is usually induced not only by the magnetic (Knight) shifts (the shift of the internal magnetic field ($H_{\mathrm{int}}$)) but also by the effect of the second order and higher order *eqQ* interactions, the increase of $\Delta H_y\,(T)$ observed here with decreasing $T$ below $T_c$ is not due to the latter origin, because the *eqQ* interaction effect is $T$-independent in the relevant temperature range. Then, the nonzero $\Delta H_y\,(T)$ observed below $T_c$ is solely due to the decrease of $H_{\mathrm{int}}$ and should be considered to be due to the reduction of the spin magnetization $M_{\mathrm{spin}}=\chi_{\mathrm{spin}} \cdot H$ (Note that the hyperfine coupling constant $A_{\mathrm{hf}}$ has been experimentally found to be positive from the $K$-$\chi$ plot.) and/or by the superconducting diamagnetism $M_{\mathrm{dia}}(<0)$. Here, we can distinguish from the field dependence of $\Delta H_y\,(T)$ if the reduction of $M_{\mathrm{spin}}$ contributes to the nonzero $\Delta H_y\,(T)$: If $M_{\mathrm{dia}}$ solely contributes to $\Delta H_y\,(T)$, $\Delta H_y\,(T)$ should decrease with increasing external field $H$, because $|\,M_{\mathrm{dia}}\,|$ decreases with $H$ in the present $H$ region ($H>>$the lower critical field $H_{c1}$). As can be seen in Fig. 2(a) at low temperatures, it clearly contradicts to the experimental observation, indicating that $M_{\mathrm{spin}}$ (and $\chi_{\mathrm{spin}}$) is reduced in the superconducting phase, as compared with that in the normal phase.

Figure 2(b) shows the $T$-dependence of $\Delta K_y(T) = f/\{\gamma_N H_{res,\,y}(T)\} - f/\{\gamma_N H_{res,\,y}(4.7\ K)\}$, where $\gamma_N$ is the nuclear gyromagnetic ratio (10.03 MHz/T for $^{59}$Co). The results seem to be naturally explained by considering that the reduction of $M_{spin}$ mainly determines the behavior of $\Delta K_y(T)$. If we neglect the contribution of $M_{dia}$ to $\Delta K_y$, the reduction of the spin component of the Knight shift, $K_{spin}$ is about 0.5 % at low temperatures in the low field region ($H$~1.5 T). Because this value of $K_{spin}$ is very close to the one (~0.45±0.10 %) estimated in the $T$ region between 120 K and 240 K by using the $K_y$-$\chi$ plot shown in Fig. 3, it is plausible that $K_{spin}$ approaches almost zero as $T$ approaches zero, suggesting that there does not exist, in the present sample, a significant fraction of normal electrons at low temperature. In this point, the sample is rather different from that used by Fujimoto *et al.*[17] in the measurement of the NMR relaxation rate $1/T_1$, where about 60 % residual density of states remains in the low temperature region (<~1.5 K). (The orbital component of the shift, $K_{orb}$ is ~2.7 %.) The inset of Fig. 3 shows the $T$ dependences of the Knight shift $K_y$ and the magnetic susceptibility $\chi$ of the powder sample. From the figure, we find that $K_y$ is almost $T$ independent below ~40 K, which is in clear contrast to the significant increase of $\chi$ with decreasing $T$. The result indicates that $\chi$ has the extrinsic component and is different from the $T$ dependence of $K_y$ reported by Michioka *et al.*[18]

We have shown detailed results of the study on the Knight shift $K_y$ , which clearly indicates that the spin susceptibility $\chi_{spin}$ is suppressed in the superconducting phase of $Na_{0.3}CoO_2{\cdot}1.3H_2O$. From the data shown in Fig. 2(b), the $T$ dependence of the upper critical field $H_{c2}$ can be determined by observing the onset temperature of the $\Delta K_y(T)$-decrease with decreasing $T$ and the results are plotted in Fig. 2(c) by the open circles together with the data reported by the resistivity measurements,[14-16] where the agreement of the two kinds of data is satisfactory. The $H_{c2}$ value extrapolated to zero temperature by using the data of the resistivity measurement, is ~8 T, which agrees well with the value ~8.2 T expected from the Pauli limit for $T_c$ of 4.5K. This supports that the superconducting pairs are in the singlet state.

We have also carried out NMR studies on several aligned crystal samples in the field $\boldsymbol{H}//c$ to exclude the possibility that the triplet spins of the Cooper pairs are pinned along the $c$ axis. Each sample has small platelets from the same batch, whose $c$ axes are aligned. The values of $T_c$ determined for many batches are found to be 3.6 ~ 4.3 K in the zero magnetic field limit. In Fig. 4(a), the spectra observed for one of the samples with $\boldsymbol{H}//c$ is shown, for example. In the spectra, there exist, besides the contribution from the superconducting bi-layer phase, those from the non-superconducting monolayer phase and the non-hydrated phase: The contributions of the non-hydrated phase indicated by the black arrows can be separated from those of the superconducting bi-layer phase and the non-superconducting monolayer one indicated by the white arrows. The existence of the monolayer phase can be recognized from the two-peak structure of the second and the third satellite line, which is known as the characteristic of the phase.[19] By using this difference of the NMR profiles between the superconducting bi-layer and non-superconducting monolayer phases with the single peak and the double peak structure, respectively,[19] we can separate the observed profile of third satellite into the two contributions as shown in the inset of Fig. 4(a) by the solid and broken lines and find that the ratio of these intensities is ~1:1. (Note that the ratio of the NMR intensities of the superconducting bi-layer phase and the monolayer one is not necessarily equal to that determined by X-ray measurements, because the latter method predominantly see the surface parts.)

The resonance frequencies $f_{res,c}$ of aligned crystals for $\boldsymbol{H}//c$ were estimated from the peak position of FT-NMR spectra of the center-line, the examples of which are shown in Fig. 4(b) at two temperatures above and below $T_c$. In Fig. 5(a), the frequency shift $\Delta f_{res,c} \equiv f_{res,c}(T)- f_{res,c}$ (~5 K) is plotted against $T$ for two values of $H$, for example. The value of $|\Delta f_{res,c}|$ at the lowest temperature studied here (1.5 K) increases with $H$, indicating that $\chi_{spin}$ is suppressed as in the case of $\boldsymbol{H}//ab$-plane, by the occurrence of the superconductivity. Figure 5(b) shows $\Delta K_c(T) \equiv f_{res,c}(T)/(\gamma_N H) - f_{res,c}(\sim 5\text{ K})/(\gamma_N H)$ against $T$. These $T$- and $H$-dependences of $\Delta K_c$ are qualitatively similar to those for $\boldsymbol{H}//ab$-plane, indicating that in the superconducting phase $\chi_{spin}$ is suppressed for $\boldsymbol{H}//c$, too.

The $K_c$ value roughly estimated from the spectra indicated in Fig. 4(a) by the white arrows for the hydrated parts of $Na_{0.3}CoO_2 \cdot 1.3H_2O$, agrees with the one estimated at 5 K by the analysis of the randomly oriented powder samples. Figure 6(a) shows the $T$-dependence of $H_{c2}$ for $\boldsymbol{H}//c$ determined from the $\Delta K_c$-$T$ curves for the four aligned crystal samples. Again the data are consistent with those determined by the resistivity measurements.[14-16]

Now, to extract the firm conclusion that the $T$- and $H$-dependences of $K_c$ exclude the possibility that the Cooper pairs have the triplet spins pinned to the $c$ axis, we discuss on the absolute value of $\Delta K_c$. Figure 6(b) shows the $T$-dependence of $\Delta K_c$ taken for the two aligned samples together with the data of $\Delta K_y$ already presented in Fig. 2 for the randomly oriented powder sample. $\Delta K_c$ exhibits only the small decrease below $T_c$ with decreasing $T$, as compared with that of $\Delta K_y$ observed for the powder sample. This can be explained by following considerations: First, we note that the spin susceptibility for $\boldsymbol{H}//c$ is known to be 1/2 of that for $\boldsymbol{H}\perp c$.[20] It is also noted that the hyperfine coupling constant $A_{hf}$ of Co nuclei with the electron spins can be considered to be nearly isotropic in the present system, because the large value of $A_{hf}$ (~ 88 kOe/$\mu_B$ as is estimated from the data shown in Fig. 3) implies that the isotropic contact interaction induced by the transferred hyperfine interaction, predominantly contributes to $A_{hf}$. Then, the anisotropy of $K_{spin}$ can be considered to be nearly equal to that of $\chi_{spin}$, though we have not taken, for the present aligned crystal sample, the data for the $K_c$-$\chi$ plot to directly estimate the value of $A_{hf}$ for $\boldsymbol{H}//c$, that is, $K_{spin}$ for $\boldsymbol{H}//c$ is expected to be about a half of $K_{spin}$ for $\boldsymbol{H}\perp c$. For this reason, the relation $|\Delta K_c| \sim (1/2)\times|\Delta K_y|$ holds. In Fig. 6(b), the magnitudes of $\Delta K_c$ expected for the bi-layer superconducting phase is as shown schematically by the broken line (Hereafter, this $\Delta K_c$ is written as $\Delta K_c^0$, and $\Delta K_c$ is used to describe the value observed for the present sample which has both the non-superconducting monolayer and the superconducting bi-layer parts. Note that the broken line tends to $\Delta K_y(T=0)\times(1/2)$, as $T$ approaches zero.)

It has to be also considered that the spectra in Fig. 4(b) contain the contribution from the non-superconducting monolayer phase, whose signal has about a half of the total intensity, as discussed above. (The contribution from the non-hydrated phase is separated from the signal of the superconducting phase.) Then, introducing a shape function, $I(\omega, T=5\text{ K})$, which describes the peak profile observed at 5 K, we use $I(\omega=2\pi f, T=2\text{ K}) \equiv (1/2)\times\{ I(\omega, T=5\text{ K}) + I(\omega+2\pi\Delta f^0_{res,c}, T=5\text{ K}) \}$ as the fitting function to the profile observed at 2 K with $\Delta f^0_{res,c}$ as the fitting parameter. (Note that the first and the second terms of the right hand side indicate the contributions from the non-superconducting monolayer and the superconducting bi-layer parts, respectively, and $\Delta f^0_{res,c}$ has a meaning of the real shift of the resonance frequency caused by the occurrence of the superconductivity in the bi-layer part. It is described by the equation $\Delta f^0_{res,c} = \gamma_N H \times \Delta K_c^0$.) By the fitting, we find, as shown in Fig. 4(b), $\Delta f^0_{res,c}$ =11 kHz. We also find from the figure that the observed peak shift $\Delta f_{res,c} = \gamma_N H \times \Delta K_c$ is about a half of $\Delta f^0_{res,c}$, that is, the observed shift $\Delta K_c$ is expected to be ~(1/2)×$\Delta K_c^0$ for the case that the signal intensity ratio of the non-superconducting monolayer and superconducting bi-layer parts is 1:1. It is shown by the dotted line in Fig. 6(b). Strictly speaking, the existence of the normal vortex cores (whose fraction is estimated to be of the order of 10 % of the total volume of the bi-layer phase) should also be considered as an origin of the $|\Delta K_c|$ reduction. In any case, the observed values of $|\Delta K_c|$ have the reasonable magnitudes, indicating that the suppression of the spin susceptibility has also been observed for $\boldsymbol{H}//c$ in the superconducting state, that is, the superconducting pairs are in the singlet state.

We also add that the deviation of the NMR peak position of the centerline of the Na nuclei, which are found by the NMR study to be located at only one crystallographic site, has hardly been observed, as shown in Fig. 6(b). It indicates that the internal field due to the superconducting diamagnetism is negligibly small. It guarantees that the observed shift of the Co NMR spectra is due to the suppression of the spin susceptibility of the conduction electrons.

Finally, a brief comment on the NQR longitudinal relaxation rate $1/T_1$ is given. In Fig. 7, the relaxation rates $1/T_1$ obtained by the present authors' group are shown for two samples of $Na_{0.3}CoO_2 \cdot 1.3H_2O$ by closed circles and closed upward triangles (for both samples $T_c \cong 4.5$ K), together with those reported in ref. 19 (open diamonds) and ref. 17 (open upward triangles). All the data in the figure seem not to have appreciable coherence peak. In the previous paper, we pointed out that a tiny coherence peak was observed by plotting the data shown here by the upward triangles in the form of $1/T_1T$ - $T$.[12] At that time, we considered that the decrease of $1/T_1$ began at a temperature slightly lower than $T_c$. However, the behavior seems not to be reproduced. We also add here the data taken for a sample of $Na_{0.3}Co_{1-x}Ir_xO_2 \cdot 1.3H_2O$ with $x$=0.005 to show how the impurity doping affects the data in the present system.

The negligibly small coherence peak of the $1/T_1T$–$T$ curve indicates, at first sight, that the superconducting order parameter has a node. The singlet pairing reported in the present paper indicates that the wave function of the Cooper pairs has an even parity. These two data might be understood by considering the *d*- or (*d*+i*d*)-wave symmetry of the order parameter. However, the small effect of non-magnetic impurities on $T_c$, which was reported in detail previously by the present authors[13,21] cannot simply be explained by such kinds of anisotropic order parameter, because non-magnetic impurities acts as the strong pair breaking centers for such the anisotropic order parameters. Now, it seems to be important to understand these results consistently. One of the possible way is, we speculate, to consider effects of the quasi particle broadening in *s*-wave superconductors: Even for the *s*-wave order parameter, the broadening caused by inelastic scatterings due to strong spin fluctuations, for example, bring about the negligibly small coherence peak, because the broadening has a significant effects on the quasi particle density of states[22] is modified significantly. Even in this case, the impurity does not act as the strong pair breaking center, whish seems to be consistent with the data reported in ref 13.

In summary, we have mainly presented the data of the NMR Knight shift taken for the field directions $\boldsymbol{H}//c$ and $\boldsymbol{H}$ within the *ab* plane, and shown that the superconducting electron pairs are in the singlet state.

This work is supported by Grants-in-Aid for Scientific Research from the Japan Society for the Promotion of Science (JSPS) and by Grants-in-Aid on priority area from the Ministry of Education, Culture, Sports, Science and Technology.

Figure caption

Fig. 1. (a) $^{59}$Co-NMR spectra of $Na_{0.3}CoO_2 \cdot 1.3H_2O$ taken at $T$= 10 K with the resonance frequency $f$= 64.000 MHz for randomly oriented powder samples. The calculated spectra are shown by the broken line. The solid and dotted lines show the calculated spectra of the central transition (-1/2 ↔ 1/2) and the first satellite transitions (-3/2 ↔ -1/2 or 1/2 ↔ 3/2), respectively. (b) and (c) Profiles of the sharp peak indicated by the arrow in (a) are shown at various temperatures for two fixed frequencies of 22.989 and 47.000 MHz, respectively.

Fig. 2. (a) Shifts of the resonance fields, $\Delta H_y(T) \equiv H_{\mathrm{res},y}(T)-H_{\mathrm{res},y}(4.7\ \mathrm{K})$ taken for $\boldsymbol{H}//y$ are plotted against $T$ for the randomly oriented powder samples at several fixed frequencies $f$. (b) Shifts of $K_y$ defined as $\Delta K_y \equiv f/\{\gamma_N H_{\mathrm{res},y}(T)\} - f/\{\gamma_N H_{\mathrm{res},y}(4.7\mathrm{K})\}$ are plotted against $T$ for the randomly oriented samples at several fixed $f$ values. (c) $T$ dependence of the upper critical field $H_{c2}$ for $\boldsymbol{H}//ab$-plane obtained from the $H$-dependence of the onset temperature of the $\Delta K_y$-decrease with decreasing $T$ is shown together with that determined by the resistivity measurements.[14-16]

Fig. 3. The $K$-$\chi$ plot obtained for the powder sample of $Na_{0.3}CoO_2 \cdot 1.3H_2O$ is shown, where the closed and open circles are the data above 100 K and below 100 K, respectively. The arrows indicate the direction of lowering $T$. The hyperfine coupling constant is estimated to be 88 kOe/$\mu_B$ and the spin and orbital components, $K_{\mathrm{spin}}$ and $K_{\mathrm{orb}}$, are estimated to be 0.45±0.1 % and 2.7±0.1 %, respectively. In the inset, the $T$ dependences of $K_y$ and the magnetic susceptibility $\chi$ are shown.

Fig. 4. (a) NMR spectra of a sample of aligned platelets taken for $\boldsymbol{H}//c$. White and black arrows indicate the signals of the hydrated phases (the superconducting bi-layer and the non-superconducting monolayer phase) and non-hydrated phase, respectively. Inset shows the profile of the third satellite with the enlarged scales, where the contributions of the superconducting bi-layer and non-superconducting monolayer phases are divided. (b) Profiles of the centerline of the spectra of the aligned platelets taken at two temperatures are shown, for example. Using the shape function $I(\omega, T$=5 K) (solid line), which describes the peak profile observed at 5 K, we introduce the function $I(\omega=2\pi f,$ $T$=2 K)≡(1/2)×{ $I(\omega, T$=5 K) + $I(\omega+2\pi\Delta f^0_{\mathrm{res},c}, T$=5 K) } and fit it to the observed profile at 2 K with $\Delta f^0_{\mathrm{res},c}$ as a fitting parameter. By the fitting, we have obtained the thick solid line for the $\Delta f^0_{\mathrm{res},c}$ value of 11 kHz. (This $\Delta f^0_{\mathrm{res},c}$ has a meaning of the actual frequency shift in the superconducting bi-layer part.) In the figure, the frequency shift $\Delta f_{\mathrm{res},c}(T) \equiv f_{\mathrm{res},c}(T)- f_{\mathrm{res},c}$ (~5 K) is shown at $T$=2 K.(The resonance frequencies $f_{\mathrm{res},c}$ are estimated from the peak position of the center-line.)

Fig. 5. (a) Observed shifts of the resonance frequencies, $\Delta f_{\mathrm{res},c}(T) \equiv f_{\mathrm{res},c}(T)- f_{\mathrm{res},c}$(~5 K) taken at two fixed fields are plotted as a function of $T$ for one of the aligned crystals. $\boldsymbol{H}//c$. (b) The data shown in (a) are plotted in the form of $\Delta K_c$ {= $f_{\mathrm{res},c}(T)/(\gamma_N H) - f_{\mathrm{res},c}$ (~5 K) $/(\gamma_N H)$}–$T$.

Fig. 6. (a) $T$ dependence of the upper critical field $H_{c2}$ for $\boldsymbol{H}//c$ estimated from the $H$-dependence of the onset temperature of the $\Delta K_c$-decrease with decreasing $T$, is shown together with the $T$ dependence of the data reported by the resistivity measurements.[15,16] (b) The $T$ dependence of the Knight shifts $\Delta K_y$ and $\Delta K_c$ observed for the field directions $\boldsymbol{H}//ab$-plane and $\boldsymbol{H}//c$, respectively, is shown. Considering the anisotropy of the spin susceptibility $\chi_{\mathrm{spin}}(\boldsymbol{H}//c)/\chi_{\mathrm{spin}}(\boldsymbol{H}\perp c)$~1/2[20] and based on the observed magnitudes of $\Delta K_y$ for $\boldsymbol{H}//ab$-plane, we roughly expect the real magnitudes $\Delta K_c^0$ for the superconducting bi-layer phase as shown by the broken line. The existence of the non-superconducting monolayer phase reduces the observed value of $\Delta K_c$ by a factor of ~1/2, as indicated by the dotted line. The $T$ dependence of $\Delta K_c$ for $^{23}$Na is also shown by the thick broken line indicated by the arrow.

Fig. 7. $T$ dependence of the nuclear relaxation rates $1/T_1$ of several samples of $Na_{0.3}CoO_2 \cdot 1.3H_2O$ is shown in the form of $\log(1/T_1)$-$\log T$, where for two sets of the data, $1/(10T_1)$ are plotted and for a set of the data, $1/(5T_1)$ are shown just for clarity. The data for a sample of $Na_{0.3}Co_{1-x}Ir_xO_2 \cdot 1.3H_2O$ with $x$=0.005 are also shown.

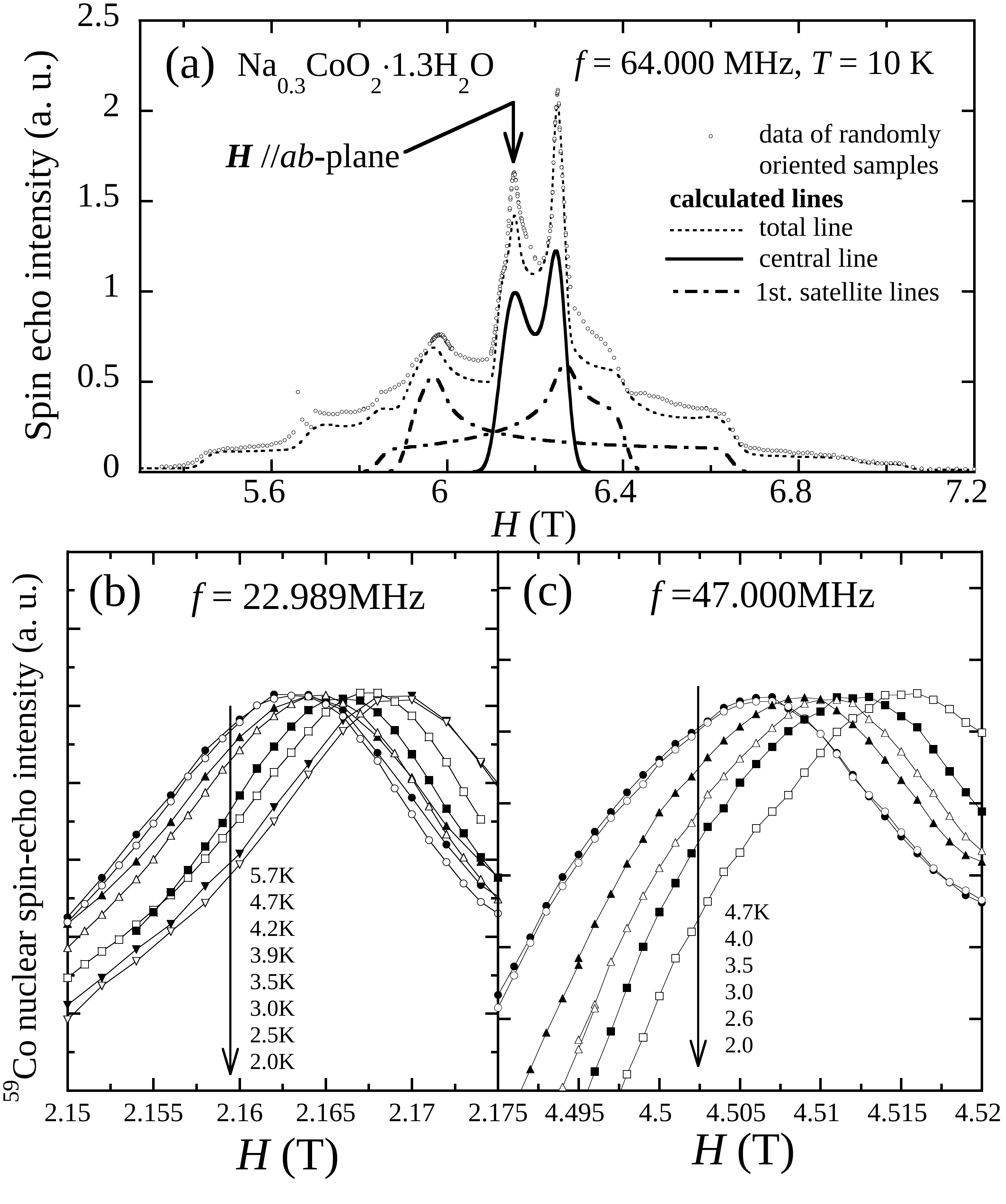

Fig. 1

Y. Kobayashi *et al.*

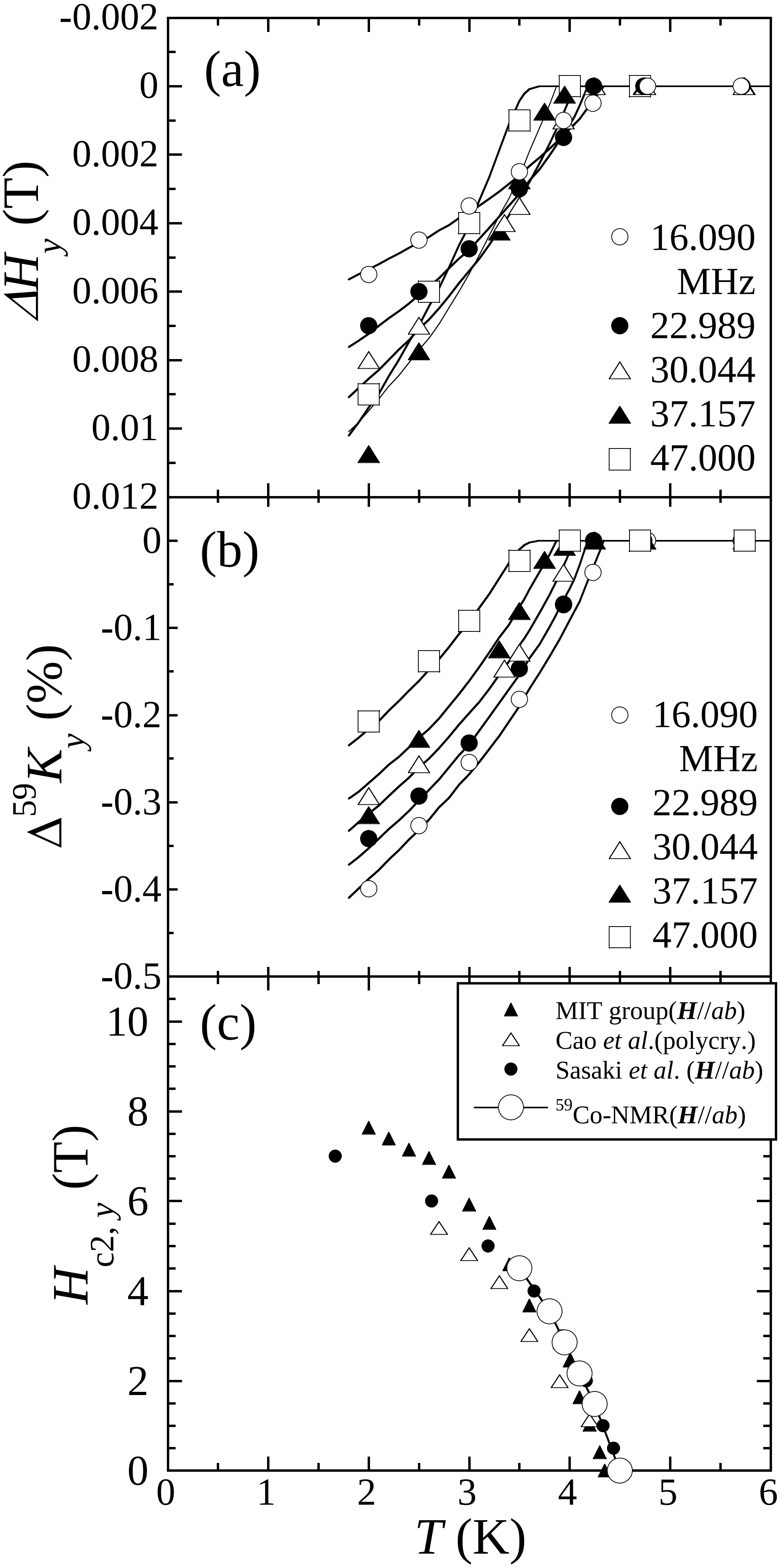

Fig. 2 Y. Kobayashi *et al.*

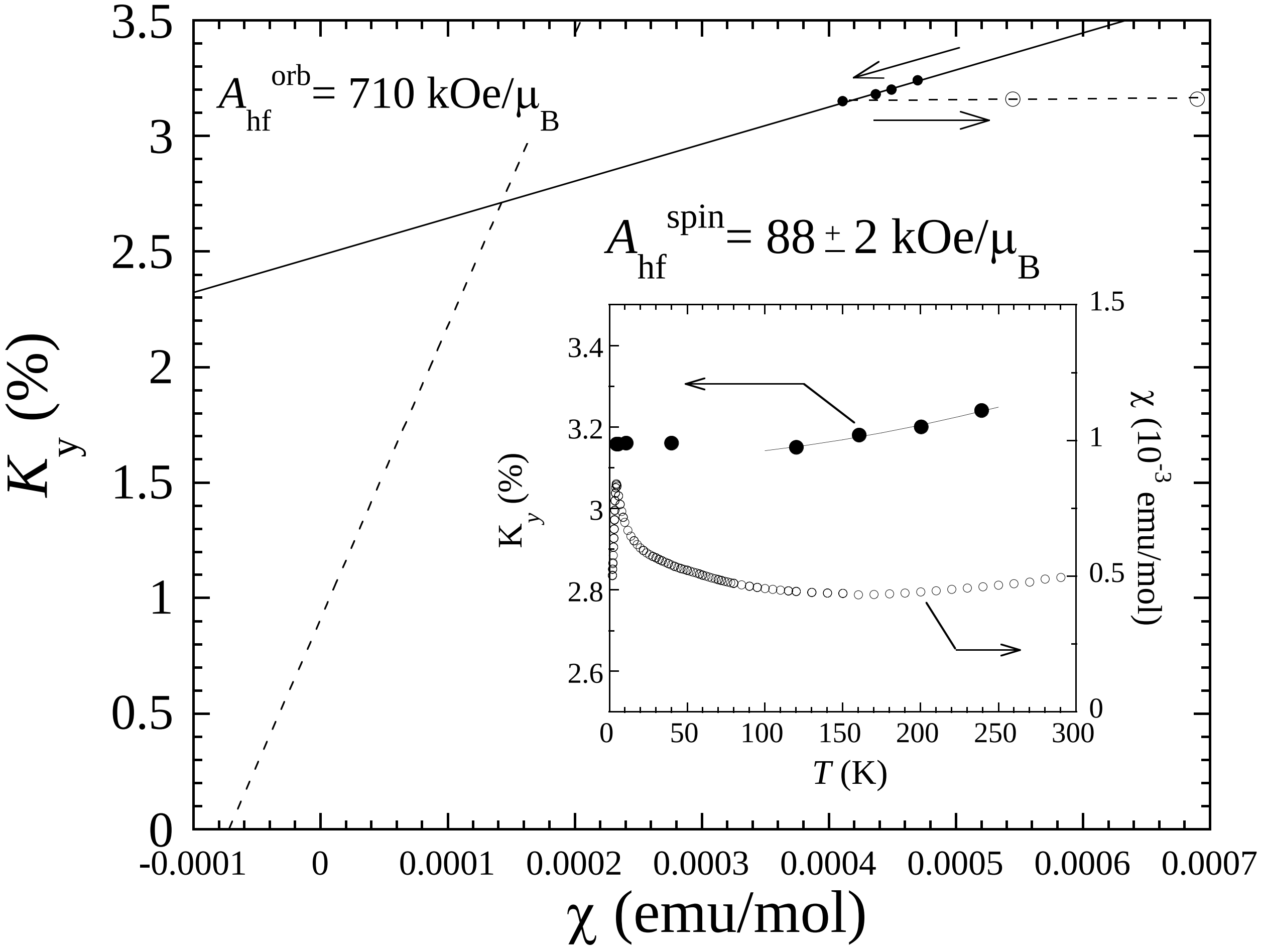

Fig. 3

Y. Kobayashi *et al.*

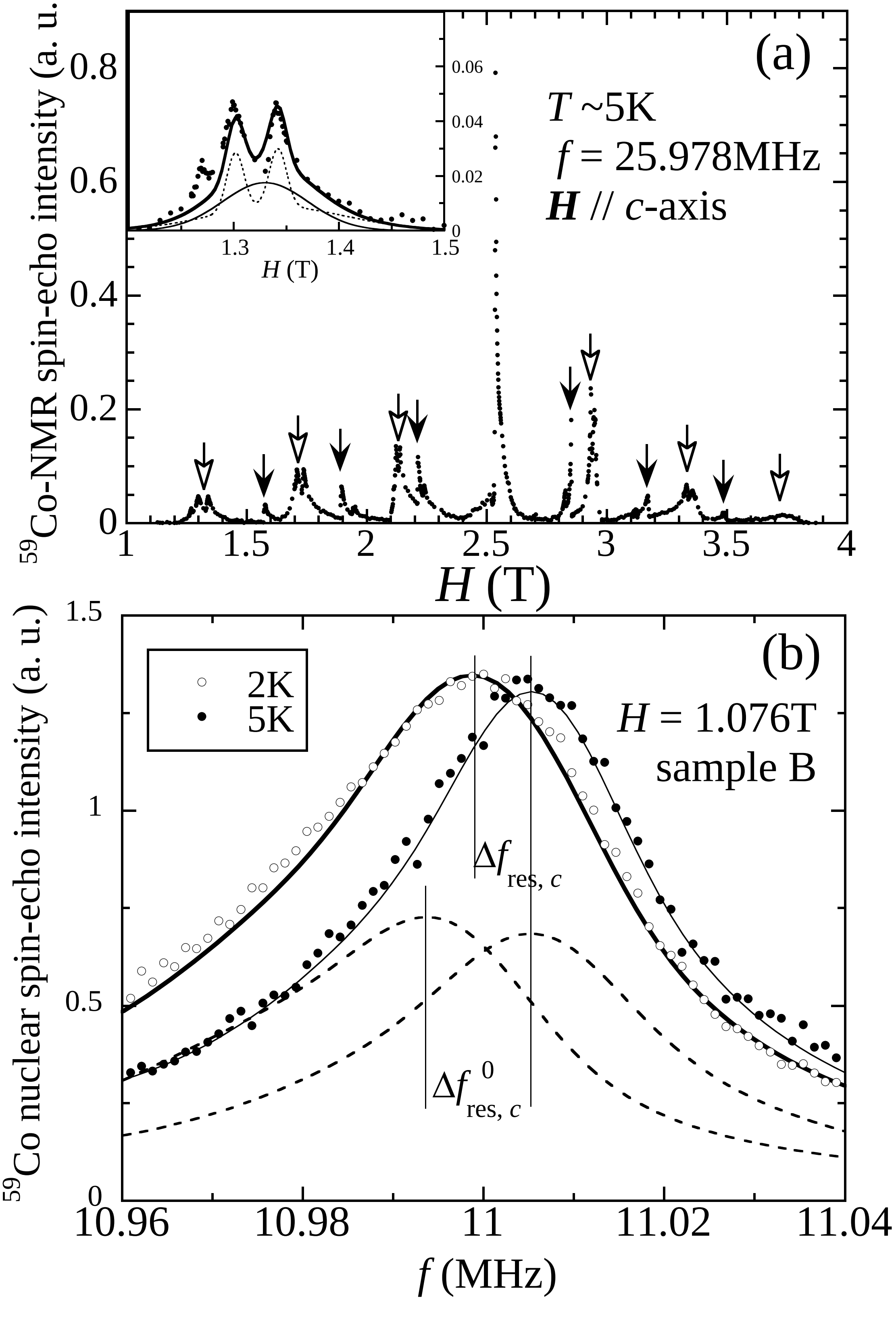

Fig. 4

Y. Kobayashi *et al.*

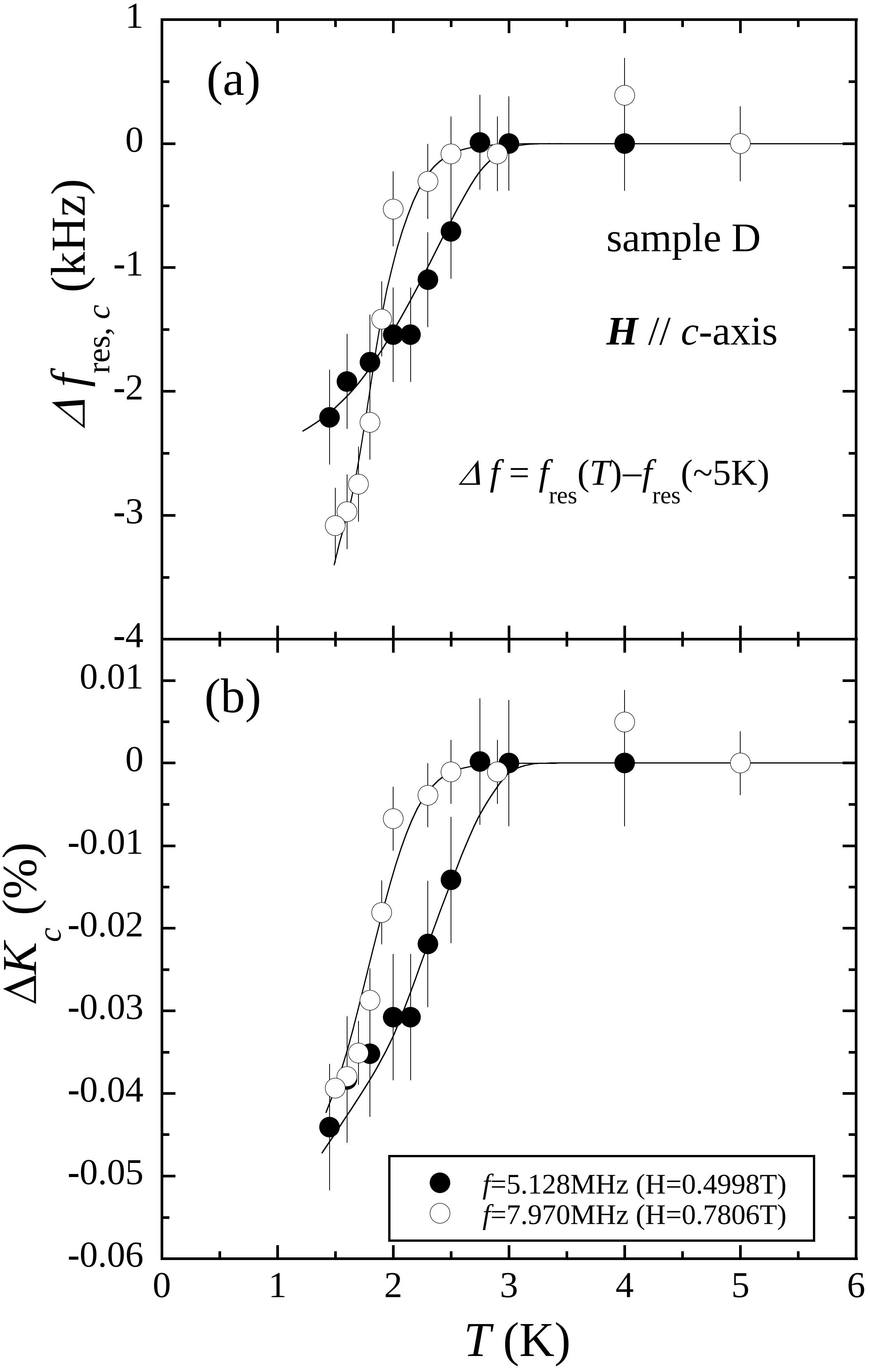

Fig. 5 Y. Kobayashi *et al.*

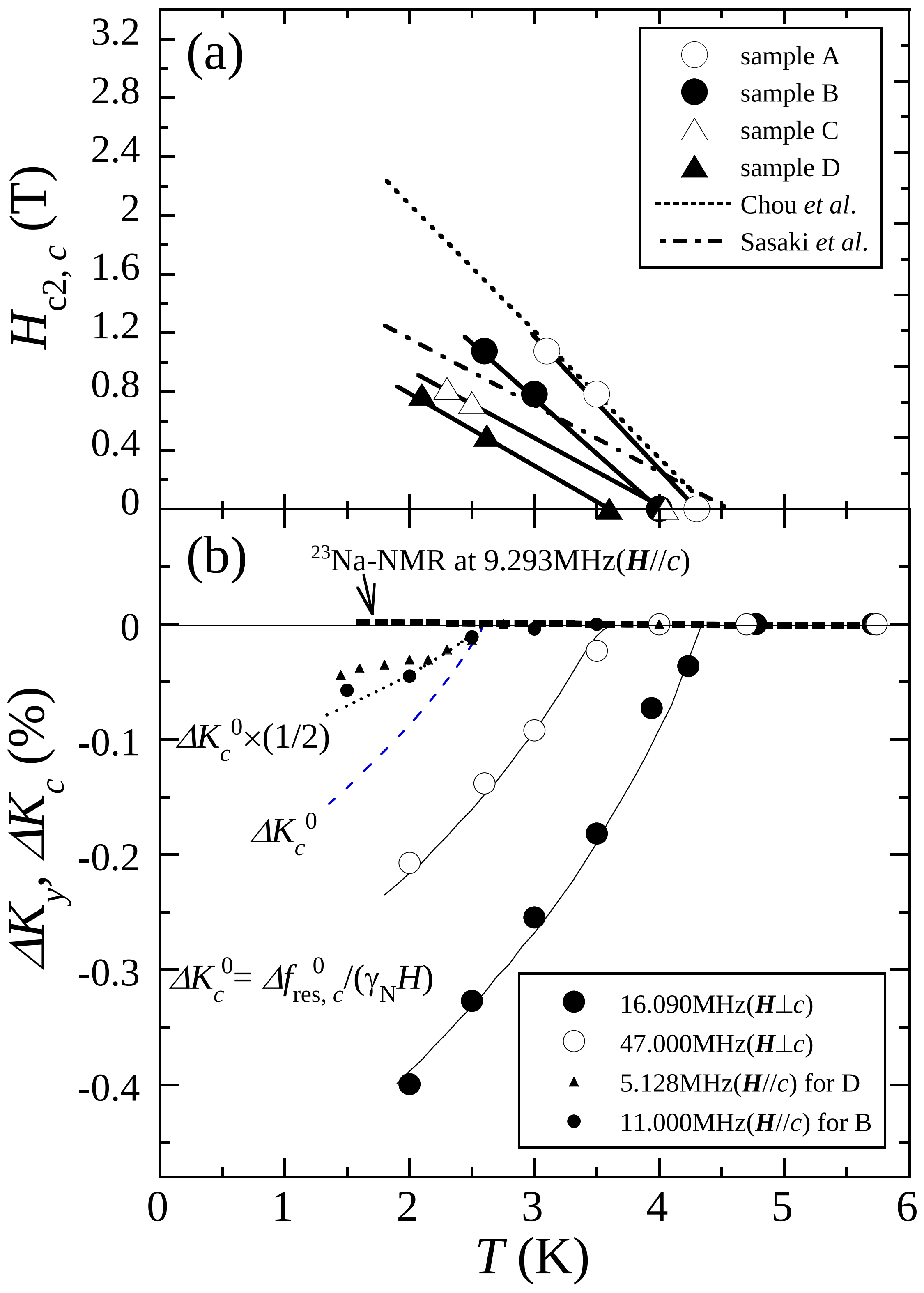

Fig. 6

Y. Kobayashi *et al.*

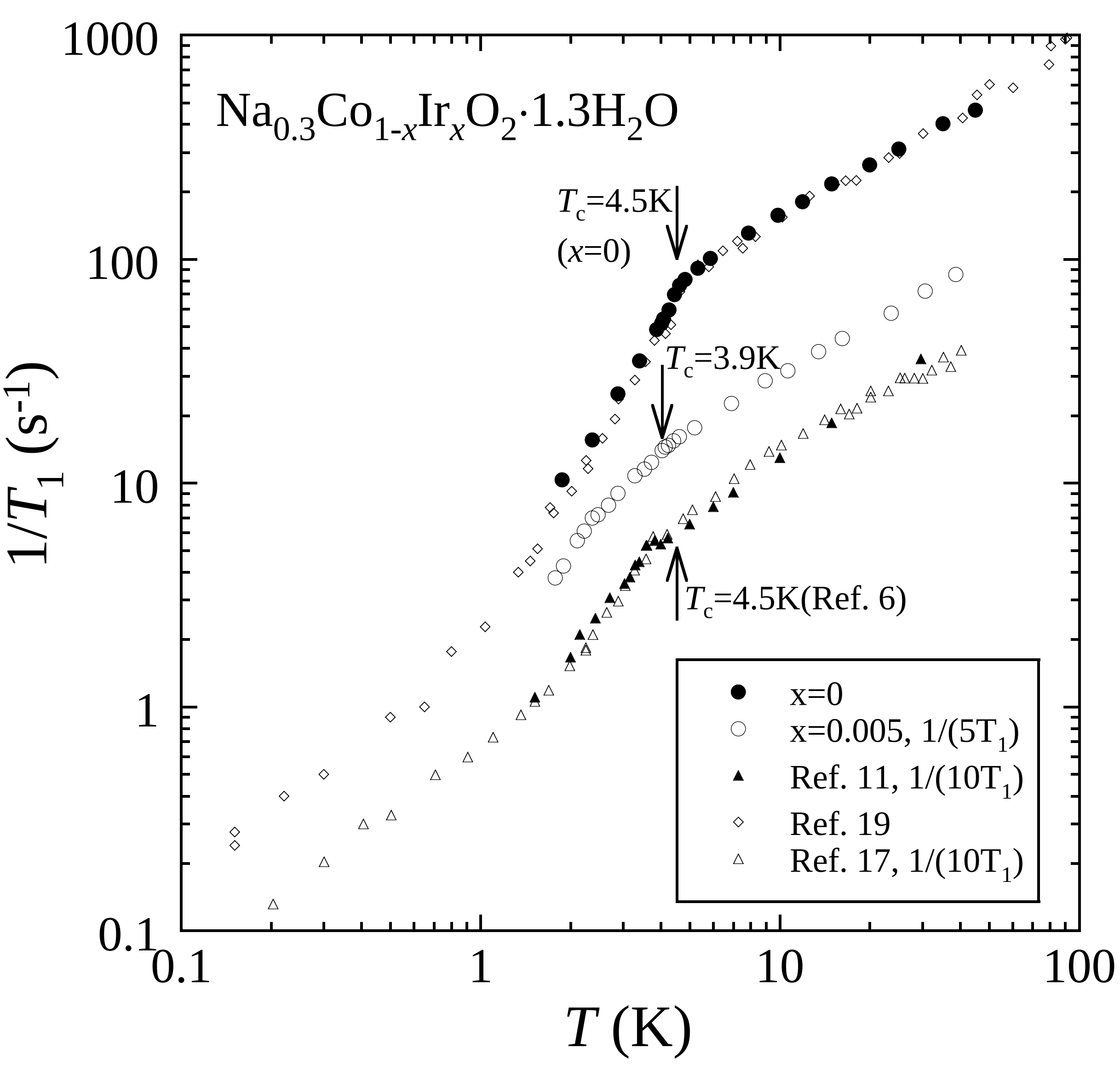

Fig. 7

Y. Kobayashi *et al.*